\documentclass[doublecol,figures]{epl2}

\usepackage{amsmath}
\usepackage{bm}
\usepackage{amsfonts}
\usepackage{amssymb}

\title{Large optical gain from four-wave mixing instabilities in semiconductor quantum wells}

\author{S. Schumacher\inst{1} \and N. H. Kwong\inst{1} \and R. Binder\inst{1}}
\shortauthor{S. Schumacher \etal}

\institute{
  \inst{1} College of Optical Sciences, University of Arizona - Tucson, Arizona 85721, USA}




\abstract{Based on a microscopic many-particle theory, we predict
large optical gain in the probe and background-free four-wave mixing
directions caused by excitonic instabilities in semiconductor
quantum wells. For a single quantum well with radiative-decay
limited dephasing in a typical pump-probe setup we discuss the
microscopic driving mechanisms and polarization and frequency
dependence of these instabilities.}

\begin{document}

\maketitle

\begin{figure}
\onefigure[scale=0.96]{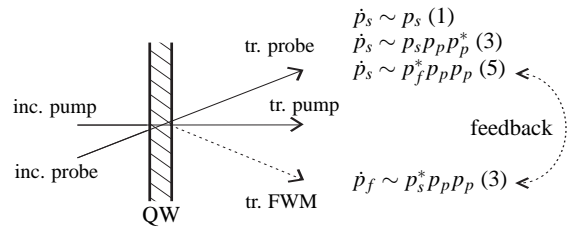}
\caption{\label{ppsetup}Illustration of the investigated pump-probe
setup, including incoming and transmitted light pulses (reflected
pulses are omitted for clarity) for probe, pump, and FWM direction
(the angle between the different propagation directions is strongly
exaggerated). Selected contributions to the probe and FWM
polarizations are schematically included in the figure. The numbers
in parentheses indicate the lowest order in the external fields in
which they appear.}
\end{figure}

In gaseous atomic or molecular systems and simple Kerr media,
four-wave mixing (FWM) processes lead, among other things, to
transverse optical instabilities (\textit{e.g.},
refs.~\cite{Yariv1977,Grynberg1988,Chang1992a,Honda1993,Firth1990}).
The interest in these FWM instabilities has recently been renewed by
the demonstration of their effectiveness for all-optical switching
at very low light intensities \cite{Dawes2005,Chen2005}. In this
Letter we argue on the basis of a microscopic many-particle analysis
that FWM instabilities can occur via nonlinear excitonic processes
in a single semiconductor quantum well (QW). {Instead of studying
spontaneous off-axis pattern formation induced by these
instabilities, we investigate their role in a pump-probe setup as
illustrated in fig.~\ref{ppsetup}. We find that the FWM
instabilities can lead to large gain in the probe and
(background-free) FWM directions that grows exponentially with the
pump pulse duration, limited by the eventual buildup of incoherent
exciton/biexciton densities.} Our analysis shows that the materials
conditions for observing these instability-induced gains, though
quite stringent, appear to be obtainable in currently available
high-quality QW samples.

FWM has been widely used in studies of microscopic processes in QWs
(\textit{e.g.},
refs.~\cite{Leo1990,Paul1996,Bartels1998,Wang1998,Cundiff1994,Banyai1995,Bolton2000,Langbein2001}).
An example of FWM driven instabilities in semiconductors is the
parametric amplification of exciton polaritons in planar QW
microcavities
\cite{Savvidis2000,Stevenson2000,Saba2001,Gippius2004,Diederichs2006,Ciuti2003,Whittaker2005,Savasta2003}.
In contrast to the microcavity, we focus on the ``simple'' system of
a single QW. Using a microscopic many-particle theory we give a
comprehensive stability analysis of the optical polarization field
induced by a normally incident pump beam, treating all vectorial
polarization state channels. In atomic systems, phase-space filling
(PSF) nonlinearities drive the instabilities, and in microcavities
it is the interaction between co-circularly (say, ++) polarized
polaritons, related to the Hartree-Fock (HF) interaction and
two-exciton (++) correlations
\cite{Ciuti2003,Whittaker2005,Savasta2003}. We find that neither PSF
nor excitonic interactions in the ++ channel are promising for
instabilities in single QWs, mainly because of excessive
excitation-induced dephasing (EID). Instead we show that
instabilities and large probe gains can be produced in single QWs by
virtual biexciton formation, which requires the pump to contain both
circular polarizations (\textit{e.g.}, to be linearly polarized).
Biexciton-induced bistabilities and instabilities have been studied
before, but in bulk semiconductors (\textit{e.g.}, for CuCl
\cite{Inoue1986,Koch1981,Peyghambarian1983,Nguyen1994}), rather than
in quasi two-dimensional QWs studied here. These early works used
bosonic models and did not consider optical gain in the off-pump
directions. It has been established (\textit{e.g.},
refs.~\cite{Axt1994a,Oestreich1995,Takayama2002}) that for typical
QW systems (\textit{e.g.}, GaAs, ZnSe) a fermionic microscopic
theory as used in this paper is needed for the understanding of
excitonic PSF, spin-dependent exciton interactions giving rise to
excitonic HF mean-field effects as well as retarded exciton
interactions that can result in the bound two-exciton state
(biexciton) and two-exciton continuum states, EID, and quantum
memory effects.


We investigate a pump-probe setup as illustrated in
fig.~\ref{ppsetup} with the light propagating in quasi-normal
incidence and with all optical pulses spectrally centered near the
1s heavy-hole (hh) exciton resonance. Assuming all other resonances
to be sufficiently far away, the coherent response of the system in
the lowest-order nonlinear regime ($\chi^{(3)}$-regime) has been
well studied. We focus our analysis on small pump intensities where
the many-particle effects listed above are dominant for the coherent
optical QW response (\textit{i.e.}, we neglect higher than
four-fermion or two-exciton correlations). The influence of
higher-order correlations and incoherent exciton contributions is
discussed later in our analysis. We start from the nonlinear
equation for the optically induced interband polarization ${p}^\pm$
($+,-$ label the circular polarization states) and perform a spatial
Fourier decomposition of ${p}^\pm$ and the exciting field $E^\pm$
with respect to the in-plane wave vector $k$. We label the Fourier
components with the subscripts $s$, $f$ and $p$ for probe (also
called signal, with $k=k_s$, assumed to be small but nonzero),
background-free FWM ($k=-k_s$), and pump ($k=0$) direction,
respectively. The resulting equations are linearized in the weak
probe field $E^\pm_s$ but solved self-consistently in the pump field
$E_p^\pm$ \cite{Buck2004,Schumacher2005a}. The equations for
$p^\pm_{s,f}$ read:
\begin{widetext}
\begin{align}\label{probemotion}
i\hbar&\dot{p}^\pm_{s,f}=(\varepsilon_x-i\gamma)p_{s,f}^\pm-\big[\phi_{1s}^{\ast}(0)-2A^{\text{PSF}}|p^\pm_p|^2\big]d_{cv}E_{s,f}^\pm
+2d_{cv}A^{\text{PSF}}\big[p_{s,f}^\pm p_p^{\pm\ast}
E_p^\pm+p_{f,s}^{\pm\ast}p_p^\pm E_p^\pm\big]  \nonumber  \\
+&2V^{\text{HF}}|p_p^{\pm}|^2p_{s,f}^\pm
+4p_p^{\pm\ast}\int_{-\infty}^{\infty}{\mathrm{d}\,t'\mathcal{G}^{\pm\pm}(t-t')p_p^\pm(t')
p_{s,f}^\pm(t')} +V^{\text{HF}}p_p^{\pm2}p_{f,s}^{\pm\ast}
+2p_{f,s}^{\pm\ast}\int_{-\infty}^{\infty}{\mathrm{d}\,t'\mathcal{G}^{\pm\pm}(t-t')p_p^\pm(t')
p_p^\pm(t')} \nonumber  \\ +&
p_p^{\mp\ast}\int_{-\infty}^{\infty}{\mathrm{d}\,t'\mathcal{G}^{\pm\mp}(t-t')\big[p_p^\mp(t')
p_{s,f}^\pm(t')+p_p^\pm(t') p_{s,f}^\mp(t')\big]} +
p_{f,s}^{\mp\ast}\int_{-\infty}^{\infty}{\mathrm{d}\,t'\mathcal{G}^{\pm\mp}(t-t')p_p^\pm(t')
p_p^\mp(t')}\,.
\end{align}
\end{widetext}
\begin{floatequation}
\mbox{\textit{see eq.~\eqref{probemotion}}}
\end{floatequation}
The nonlinear pump equation for $p_p^\pm$ (not shown) involves the
same nonlinear processes. We note that the solution to
eq.~(\ref{probemotion}) goes beyond the $\chi^{(3)}$-limit and
includes the pump polarization up to arbitrary order. Here,
$\varepsilon_x$ is the 1s-hh exciton energy, $\gamma$ a
phenomenological excitonic dephasing constant (not including the
radiative decay), $d_{cv}$ the interband dipole matrix element,
$\phi_{1s}(\mathbf{r})$ the two-dimensional exciton wavefunction,
$A^{\text{PSF}}$ the excitonic PSF constant, and $V^{\text{HF}}$ the
excitonic HF matrix element. Unless otherwise noted, the time
argument is $t$.  The correlation kernels $\mathcal{G}$ are given by
$\mathcal{G}^{++}=\mathcal{G}^{--}=\tilde{G}^+$ and
$\mathcal{G}^{+-}=\mathcal{G}^{-+}=\tilde{G}^++\tilde{G}^-$, with
$\tilde{G}^{\pm}$ as defined in eq.~(22) in
ref.~\cite{Takayama2002}, including a two-exciton dephasing rate
$2\gamma$. The propagation of the optical field $E^\pm$ across the
QW is described with a transfer-matrix method accounting for
radiative corrections/decay and assuming the QW to be infinitely
thin (\textit{e.g.}, ref.~\cite{Khitrova1999}).

\begin{figure}
\begin{center}
\onefigure[scale=0.8]{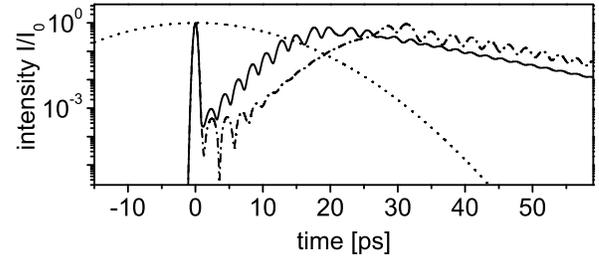}
\end{center}
\caption{\label{threshold} Transmitted intensity of a
$500\,\mathrm{fs}$ probe (normalized to the peak intensity $I_0$ of
the incoming probe). The shape of the $20\,\mathrm{ps}$ pump pulse
is included as the dotted line. Results are shown for XX pump-probe
polarization, zero pump-probe delay, pump peak intensities
$22\,\mathrm{kWcm^{-2}}$ (solid) and $5.3\,\mathrm{kWcm^{-2}}$
(dashed-dotted), and equal pump and probe central frequencies tuned
to $-0.6\,\mathrm{meV}$ (solid) and $-1.2\,\mathrm{meV}$
(dashed-dotted) below the exciton resonance, respectively.}
\end{figure}

From a time-domain solution\footnote{Parameters:
$\varepsilon_x=1.4965\,\mathrm{eV}$,
$E_b^{xx}\approx1.84\,\mathrm{meV}$, $d_{cv}=4\,e\AA$,
$\gamma=0.01\,\mathrm{meV}$, $A^{\text{PSF}}=4a_0^x\sqrt{2\pi}/7$,
$V^{\text{HF}}=2\pi(1-315\pi^2/4096)a_0^{x2}E_b^x$, with
$a_0^x\approx170\,\AA$, $E_b^x\approx13\,\mathrm{meV}$.} of
eq.~(\ref{probemotion}) and the equation for the nonlinear pump
polarization, we find for co-linear (XX) pump-probe polarization
configuration for certain pump detunings
$\Delta\varepsilon=\hbar\omega_{p}-\varepsilon_{x}$ and above a
certain pump threshold intensity $I_{\text{th}}$ a large energy
transfer from the pump into the probe and FWM directions. To
illustrate this phenomenon, fig.~\ref{threshold} shows  the
transmitted probe intensity for two different $\Delta\varepsilon$
which -- after the initial probe has passed -- grows (almost)
exponentially until the pump-induced QW polarization (not shown)
falls below a certain threshold value (calculations for a
high-quality GaAs QW at low temperatures where the radiative decay
dominates the homogeneous linewidth as, \textit{e.g.}, in
ref.~\cite{Langbein2001}). For the results in fig.~\ref{threshold}
about two orders of magnitude of time-integrated gain in the probe
direction is found. We find similar outgoing fluences in the
background-free FWM directions (not shown).

To analyze the gain mechanism, we supplement the full time-dependent
solution by a linear stability analysis (LSA). The LSA is done
without an incoming probe field and for a monochromatic cw pump
field $E^\pm_{p}(t)=\tilde{E}^\pm_{p}\mathrm{e}^{-i\omega_pt}$ and
pump polarization
$p^\pm_{p}(t)=\tilde{p}^\pm_{p}\mathrm{e}^{-i\omega_pt}$, with
$\dot{\tilde{p}}^\pm_p=\dot{\tilde{E}}^\pm_p=0$ ($\omega_p$ is the
pump frequency). We evaluate the memory integrals in
eq.~(\ref{probemotion}) in Markov approximation [$p_{s,f}(t')\approx
p_{s,f}(t)\mathrm{e}^{i\omega_p(t-t')}$] for the two-exciton
continuum in the correlation kernels $\mathcal{G}^{\pm\pm}$,
$\mathcal{G}^{\pm\mp}$. In contrast, exact treatment of the bound
biexciton state [included in the correlation kernel
$\mathcal{G}^{\pm\mp}$ in eq.~(\ref{probemotion})
\cite{Takayama2002}] via the time-dependent amplitudes $b_{s,f}(t)$
is crucial to include the related quantum memory effects
(\textit{e.g.}, refs.~\cite{Axt2004,Shahbazyan2000,Sieh1999}) which
-- as our analysis reveals -- give rise to additional unstable
modes. According to eq.~(\ref{probemotion}) the term driving the
bound biexciton amplitudes $b_{s,f}(t)$ is proportional to $p_p^\mp
p_{s,f}^\pm+p_p^\pm p_{s,f}^\mp$. With the ansatz
$p_{s,f}(t)=\tilde{p}_{s,f}(t)\mathrm{e}^{-i\omega_pt}$ and
$b_{s,f}(t)=\tilde{b}_{s,f}(t)\mathrm{e}^{-i2\omega_pt}$ the probe
and FWM dynamics take the form
$\frac{\mathrm{d}}{\mathrm{d}t}{\tilde{\mathbf{p}}}(t)=M\,{\tilde{\mathbf{p}}}(t)$,
with ${\tilde{\mathbf{p}}}(t)=\big[{\tilde{p}}^+_s(t),
{\tilde{p}}_f^{+\ast}(t), {\tilde{p}}^-_s(t),
{\tilde{p}}_f^{-\ast}(t), \tilde{b}_s(t),
\tilde{b}_f^{\ast}(t)\big]^T$ where the matrix $M$ follows from
eq.~(\ref{probemotion}). If at least one of the eigenvalues
$\lambda_{i}$ of $M$ fulfills $\mathrm{Re}\{\lambda_{i}\}>0$, the
system is unstable. {An arbitrarily small seed of $p^\pm_{s,f}$
would grow exponentially, until the matrix $M$ ceases to describe
the system correctly.}

On the PSF-level [only line 1 of eq.~(\ref{probemotion})] the
nonlinear terms from left to right are: (i) non-grating
Pauli-blocking, (ii) self wave mixing (SWM) including power
broadening with the general structure $i\dot{p}_{s,f}\sim
p_{s,f}p_p^\ast E_p $, and (iii) the phase-conjugate oscillation
(PCO) with the structure $i\dot{p}_{s,f}\sim p_{f,s}^\ast p_pE_p$
which gives the feedback between probe and FWM signal (compare
fig.~\ref{ppsetup}) which drives the instability in atomic systems,
\textit{e.g.}, in ref.~\cite{Yariv1977}. Going beyond the PSF-level,
in the second line of eq.~(\ref{probemotion}), HF exciton Coulomb
interaction and excitonic correlations [the two-exciton continuum in
the electron-spin triplet ($++$) channel] contribute to the
two-exciton interaction. While the first two of these terms are of
the SWM type ($i\dot{p}_{s,f}\sim p_{s,f}p_p^\ast p_p $) and
basically blue-shift the exciton resonance in
eq.~(\ref{probemotion}) and lead to additional EID, the last two
terms have the general structure of a PCO contribution
($i\dot{p}_{s,f}\sim p_{f,s}^\ast p_p p_p $) mediated by the
two-exciton Coulomb interaction. The third line in
eq.~(\ref{probemotion}) contains the Coulomb interaction in the
two-exciton electron-spin singlet ($+-$) channel, including both the
bound biexciton state and the two-exciton continuum in this channel,
where the first term is of the SWM type and the last term has the
structure of a PCO coupling.

We start the discussion with a $\sigma^+$ polarized pump. Without
Coulomb interaction, only the PSF-PCO can lead to an instability as
it is the case in atomic systems. However, within our model with
zero in-plane exciton polariton dispersion, one can show
analytically that PSF alone does not yield an instability (i.e., all
$\mathrm{Re}\{\lambda_{i}\}<0$), which is analogous to the forward
propagating case in atomic systems \cite{Grynberg1988}; PSF-SWM and
PSF-PCO counteract each other. We note that an extension of
Eq.~(\ref{probemotion}) to realistic nearly quadratic dispersions
\cite{Tassone1990} [$\varepsilon_x \equiv \varepsilon_x(k)$] can --
in principle -- yield instabilities for certain positive pump
detuning $\varepsilon_x(0)<\hbar\omega_p<\varepsilon_x(k_s)$,
however our numerical studies have shown that they occur only at
extremely high intensities, rendering the instabilities unlikely to
be experimentally observable and justifying the neglect of the
in-plane dispersion.

Without in-plane dispersion and including only HF terms (which,
physically, leads to an exciton blue-shift caused by the HF-SWM
terms), we found from a numerical diagonalization of $M$
instabilities in a small spectral window for positive pump detuning,
mediated by the HF-PCO contribution. In planar semiconductor
microcavities, the parametric polariton amplification has been
attributed to this instability
\cite{Savvidis2000,Stevenson2000,Saba2001,Ciuti2003,Gippius2004,
Whittaker2005,Diederichs2006}. However, two-exciton correlations in
the electron-spin triplet ($++$) channel \cite{Takayama2002} have
been found to weaken this instability \cite{Savasta2003}, and in our
case of a single QW we find that for realistic parameters the
instability is inhibited when these correlations are taken into
account; mainly because of EID due to two-exciton scattering.

\begin{figure}
\begin{center}
\onefigure[scale=0.8]{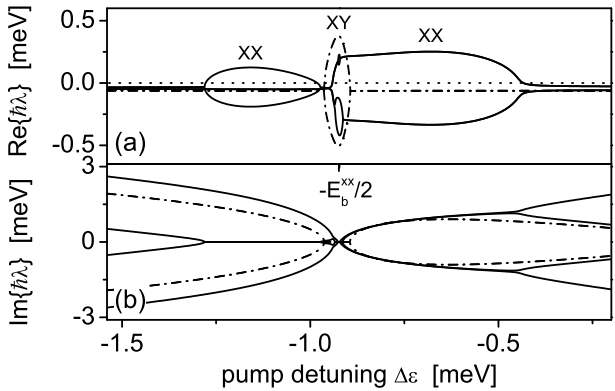}
\end{center}
\caption{\label{eigenvalues.fig} Linear stability analysis for a
linearly polarized pump for steady-state total coherent exciton
density $n_x^{\text{total}}=1.6\times10^{10}\mathrm{cm^{-2}}$. Shown
are the real parts (a) and imaginary parts (b) of the eigenvalues
$\lambda_{i}$ of the matrix $M$ for negative pump detuning.
{Eigenvalues are represented by solid lines for the co-linear (XX)
and by dashed-dotted lines for the cross-linear (XY) polarization
configuration. The corresponding pump intensity and exciton
($|p^X_p|^2$) and biexciton ($2|b_p|^2$) contributions to
$n_x^{\text{total}}$ are depicted in fig.~\ref{density.fig}. The
dotted line in panel (a) separates the stable
($\mathrm{Re}\{\lambda\}<0$) from the unstable
($\mathrm{Re}\{\lambda\}>0$) regime.}}
\end{figure}

In contrast, a regime where excitonic correlations are strong but
EID is kept small can be accessed for negative pump detuning
$\Delta\varepsilon<0$ with a linearly (say X) polarized pump that
also excites the excitonic correlations in the electron-spin singlet
($+-$) channel. The corresponding eigenvalues of $M$ are shown in
fig.~\ref{eigenvalues.fig} for a fixed total coherent exciton
density defined as $n_x^{\text{total}}\equiv|p^X_p|^2+2|b_p|^2$ with
the coherent exciton ($|p^X_p|^2$) and bound biexciton ($|b_p|^2$)
densities (for $\Delta\varepsilon<0$ we have checked that density
contributions from the two-exciton continuum are negligible compared
to $n_x^{\text{total}}$). The pump polarization $p^X_p$ induced by
the X polarized pump follows from the (cubic) nonlinear pump
equation and $b_p$ is the bound biexciton amplitude driven by a
source term proportional to $p_p^+p_p^-$.

\begin{figure}
\begin{center}
\onefigure[scale=0.77]{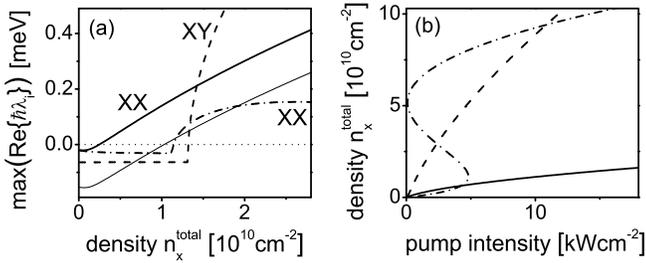}
\end{center}
\caption{\label{detunings.fig} (a) Real part of the eigenvalue
$\lambda_i$ of $M$ with the largest real part corresponding to
co-linear (XX) and cross-linear (XY) pump-probe polarization
configuration, respectively. (b) Total coherent exciton density
$n_x^{\text{total}}$ vs. pump intensity inside the QW. (a), (b)
Results are shown for different pump detunings,
$\Delta\varepsilon=-0.6\,\mathrm{meV}$ (solid),
$\Delta\varepsilon=-0.92\,\mathrm{meV}$ (dashed), and
$\Delta\varepsilon=-1.2\,\mathrm{meV}$ (dashed-dotted). The thin
solid line in (a) shows the same result as the solid line but for a
ten times larger dephasing $\gamma=0.1\,\mathrm{meV}$.}
\end{figure}

In fig.~\ref{eigenvalues.fig}(a) we find three different unstable
regions  ($\mathrm{Re}\{\lambda\}>0$) caused by the biexcitonic
($+-$) PCO terms in eq.~(\ref{probemotion}). The labels XX and XY
denote the vectorial polarizations of the pump (always X) and the
unstable modes or probe fluctuations (either X or Y).  The
dependence of the eigenvalues with the largest real parts (which,
when positive, are the maximum growth rates) on $n_x^{\text{total}}$
is shown in fig.~\ref{detunings.fig}(a) [with
fig.~\ref{detunings.fig}(b) relating  $n_x^{\text{total}}$ to the
pump intensity]. Clearly, the LSA helps us understand the temporal
growth in fig.~\ref{threshold}, with maximum coherent densities
$n_x^{\text{total}}\approx1.8\times10^{10}\,\mathrm{cm^{-2}}$
($22\,\mathrm{kWcm^{-2}}$) and
$\approx3.5\times10^{10}\,\mathrm{cm^{-2}}$
($5.3\,\mathrm{kWcm^{-2}}$).

\begin{figure}
\begin{center}
\onefigure[scale=0.9]{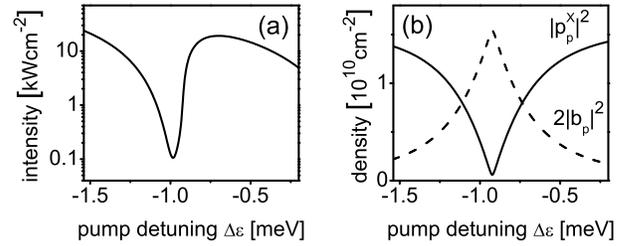}
\end{center}
\caption{\label{density.fig} {(a) Pump intensity corresponding to
the total coherent exciton density
$n_x^{\text{total}}=1.6\times10^{10}\mathrm{cm^{-2}}$ used in
fig.~\ref{eigenvalues.fig}. (b) Exciton ($|p^X_p|^2$) and biexciton
($2|b_p|^2$) contributions to $n_x^{\text{total}}$.}}
\end{figure}

We consider a range of relatively small total coherent densities
$n_x^{\text{total}}\lesssim3\times10^{10}\,\mathrm{cm^{-2}}$ to
minimize the influence of higher-order correlations not included in
our theory. This restriction limits the maximum growth rates in the
unstable regime [fig.~\ref{detunings.fig}(a)] and therefore dictates
the minimum timescale that is required to achieve a desired amount
of gain. In principle, accumulation of an incoherent
exciton/biexciton density in the system on this timescale (leading
to additional EID) can cap the possible amount of gain. However, for
the time scales in fig.~\ref{threshold} we believe that additional
EID from incoherent densities would not crucially affect the
presented results.\footnote{For pump excitation below the exciton
resonance, energy conservation dictates that only scatterings from
coherent and incoherent bound biexcitons (the incoherent biexciton
density does not exceed $\approx2.5\times10^{9}\,\mathrm{cm^{-2}}$
for the results shown in fig.~\ref{threshold}) can efficiently cause
EID losses for the coherent exciton and biexciton densities.
Assuming that the scattering rates for processes involving at least
one bound biexciton are considerably lower than for two-exciton
scattering, these EID contributions are negligible for the time
scales considered in this paper.} {In addition to the EID from
incoherent densities, the exponential growth of gain with increasing
pump length can also be limited by depletion of the pump signal, if,
after a certain growth period, the probe and FWM signals have become
sufficiently strong so that linearization of the polarization
dynamics in these signals is no longer valid.} Finally, in
fig.~\ref{detunings.fig}(a) we show how the instability threshold
density changes by increasing the effective dephasing in the system
(\textit{e.g.}, by increasing temperature or due to EID). As a
relatively sharp spectral feature around the two-photon biexciton
resonance [fig.~\ref{eigenvalues.fig}(a)] the XY instability may not
be robust against even small deviations from the model conditions
considered here. Therefore, in the following, we focus our attention
on the unstable modes in the XX configuration.

We can distinguish (i) an XX instability below
($\Delta\varepsilon<-E_b^{xx}/2$) and (ii) above
($\Delta\varepsilon>-E_b^{xx}/2$) the two-photon biexciton resonance
at $\Delta\varepsilon=-E_b^{xx}/2\approx-0.92\,\mathrm{meV}$.
Focusing on (i), we see from fig.~\ref{detunings.fig}(b) that it
occurs in a region of bistability (\textit{i.e.}, three possible
steady state coherent exciton densities for fixed intensity). {In
the corresponding time-domain calculation in fig.~\ref{threshold}
the instability of the intermediate branch of the bistable pump
solution (cf. ref.~\cite{Nguyen1994}) is reflected in the
exponentially growing probe signal over time, and as demonstrated in
fig.~\ref{threshold} can give rise to large probe gain. However, we
note that in this bistable regime the system dynamics sensitively
depends on the `initial conditions', determined by the overall
temporal pump pulse shape. In fig.~\ref{threshold} we use a Gaussian
pump pulse with intensity and pulse length chosen such that the
pulse induces a coherent density on the intermediate (unstable)
branch in the bistable regime [cf. dashed-dotted line in
fig.\ref{detunings.fig}(b)]. However, for a different choice of pump
parameters the induced polarization may have mainly contributions
from the lower and upper (stable) branches, so that no exponential
growth (instability) in fig.~\ref{threshold} would be observed.}
Turning to the instability (ii), we find a much lower threshold (as
low as $n_x^{\text{total}}\approx3\times10^{9}\mathrm{cm^{-2}}$)
than for the other configurations. Furthermore,
fig.~\ref{eigenvalues.fig}(b) explains the apparent beating on the
probe signal (solid line in fig.~\ref{threshold}) during the
exponential growth period. In case (ii), the unstable modes exist in
pairs with complex conjugate eigenvalues. Whereas the growth rates,
$\mathrm{Re}\{\lambda\}$, are the same for the two modes in each
pair, the frequencies, $\mathrm{Im}\{\lambda\}$, have opposite
signs. The unstable growth of the probe and {FWM} signals occurs at
the two frequencies $\omega_p\pm|\mathrm{Im}\{\lambda\}|$. This is
inherently related to a non-Markovian quantum memory effect, brought
about by the time-retarded structure of the biexcitonic SWM term in
eq.~(\ref{probemotion}).

Our analysis is applicable to other semiconductor materials provided
suitable scalings in lengths (with $a_0^x$) and energies (with
$E_b^x$) are made. Since we believe that FWM instabilities in QWs
can be expected as long as the excitonic dephasing is not much
larger than the radiative decay (\textit{e.g.},
$\approx0.05\,\mathrm{meV}$ for GaAs \cite{Langbein2001},
$\approx0.3\,\mathrm{meV}$ for ZnSe), materials with a very small
ratio $\gamma / E^x_b$ would be ideal.

In conclusion, we predict optical FWM instabilities and
corresponding strong probe and FWM signal gain in a semiconductor
QW. Further research, \textit{e.g.}, into possible transverse
pattern formation and propagation effects in multiple QWs, would be
desirable.

\acknowledgments {We thank Arthur L. Smirl, University of Iowa, for
generously sharing his ideas and insights on the subject matter with
us.} This work has been supported by ONR, DARPA, and JSOP. \mbox{S.
Schumacher} gratefully acknowledges support by the Deutsche
Forschungsgemeinschaft (SCHU~1980/3-1).

\bibliography{./instEPLlib}
\bibliographystyle{eplbib}

\end{document}